\documentclass[prl,twocolumn,showkeys,showpacs,superscriptaddress,10pt
]{revtex4-1}
\usepackage{subfigure}
\usepackage{graphicx}  
\usepackage{grffile}
\usepackage{amsthm}
\usepackage{epsfig}
\usepackage{amsmath,amssymb,amsfonts,mathtools,bbm,MnSymbol,mathrsfs,xfrac}   
\usepackage{longtable} 
\usepackage{tabularx}
\usepackage[breaklinks=true,colorlinks=true,linkcolor=blue,urlcolor=blue,citecolor=blue]{hyperref}
\usepackage{comment}
\usepackage{color}
\usepackage[makeroom]{cancel}

\usepackage{soul}

\renewcommand{\[}{\begin{equation}}
\renewcommand{\]}{\end{equation}}

\newcommand{\pro}[2]{|#1\rangle\langle#2|}

\newcommand{\tr}{\mathrm{tr}}

\newcommand{\R}{{\hat{\rho}}}

\newcommand{\C}{{\mathcal{C}}}
\renewcommand{\P}{\hat{P}}

\newcommand{\HS}{\mathcal{H}}

\definecolor{mygray}{gray}{0.6}

\theoremstyle{definition}
\newtheorem{definition}{Definition}
\newtheorem{theorem}{Theorem}

\begin{document}

\title{Quantum coarse-grained entropy and thermodynamics}

\author{Dominik \v{S}afr\'{a}nek}
\email{dsafrane@ucsc.edu}
\affiliation{SCIPP and Department of Physics, University of California, Santa Cruz, CA 95064, USA}
\author{J. M. Deutsch}
\affiliation{Department of Physics, University of California, Santa Cruz, CA 95064, USA}
\author{Anthony Aguirre}
\affiliation{SCIPP and Department of Physics, University of California, Santa Cruz, CA 95064, USA}

\date{\today}

\begin{abstract}
We extend classical coarse-grained entropy, commonly used in many branches of physics, to the quantum realm. We find two coarse-grainings, one using measurements of local particle numbers and then total energy, and the second using local energy measurements, which lead to an entropy that is defined outside of equilibrium, is in accord with the thermodynamic entropy for equilibrium systems, and reaches the thermodynamic entropy in the long-time limit, even in genuinely isolated quantum systems. This answers the long-standing conceptual problem, as to which entropy is relevant for the formulation of the second thermodynamic law in closed quantum systems. This entropy could be in principle measured, especially now that experiments on such systems are becoming feasible.
\end{abstract}

\maketitle

Entropy, and its increase, are crucial concepts applied across an array of physical theories and systems. Yet entropy has many distinct proposed definitions~\cite{balian2004entropy,goold2016role}, and there are situations in which it is unclear which if any of these definitions apply, and which entropies can be considered as suitable candidates for entropy appearing in the second thermodynamic law.

Consider a closed physical system -- be it isolated in a laboratory or ``the whole Universe'' -- undergoing Hamiltonian evolution with no interaction with the outside world.  Classically, thermal entropy and its increase are generally treated through coarse-graining of phase space. A system can have time-evolving coarse-grained quantities, such as order parameters, energy, and currents -- as is the case in fluid dynamics or phase transitions~\cite{ma1985statistical,sethna2006statistical}. An entropy measure can then naturally and generically rise if it attributes higher entropy to coarse-grained states of greater phase-space volume.

In other words, coarse-graining describes the macroscopic degrees of freedom, and the second thermodynamic law can be viewed as the tendency of the microscopic state of the system to naturally evolve into a macroscopic state of larger phase-space volume. The second thermodynamic law that ``Total entropy of an isolated system cannot decrease over time'' then follows easily when applied to this ``Boltzmann'' entropy -- even if the {\em Gibbs} entropy is conserved, or zero.

In a closed quantum system, the standard von Neumann entropy is constant (and zero for a pure state), in close correspondence to the classical Gibbs entropy. Such entropy therefore cannot underlie the second thermodynamic law. The natural question to ask is then: What kind of entropy \emph{does} increase in an isolated quantum system? In analogy with classical thermodynamics, to find such entropy it would be desirable to define a notion of coarse-graining, and with it a quantum equivalent of Boltzmann entropy. This has not, we would argue, previously been done in any natural or compelling way. This is because the corresponding procedure of coarse-graining has been hard to formulate in quantum mechanics due to the lack of commutation of conjugate degrees of freedom. This problem is particularly severe when discussing coarse-grained entropy, where phase space volume is a crucial concept.

Instead, other quantum mechanical definitions of entropy have been developed, such as diagonal entropy~\cite{tolman1938principles,ter1954elements,jaynes1957information2,polkovnikov2011microscopic}, entropy of an observable~\cite{ingarden1962quantum,grabowski1977continuity,anza2017information}, entanglement entropy~\cite{bombelli1986quantum,srednicki1993entropy,bennet1996concentrating,bennet1998quantum}, that can give rise to the thermodynamic entropy even in pure states~\cite{polkovnikov2011microscopic,ikeda2015second,goldstein2013second,deutsch2010thermodynamic,deutsch2013microscopic,santos2012weak}, and information-theoretic quantities such as quantum relative entropy~\cite{sagawa2012nonequilibrium,modi2010unified,deffner2010generalized}, and max-entropy~\cite{konig2009operational,del2011thermodynamic}. However, their relation to the coarse-graining used in classical systems is obscure or lacking, and they can behave oddly in certain cases.

In this letter, we argue that we {\em can}, in fact, define a coarse-graining in quantum mechanics in a satisfactory and surprisingly elegant way. The resulting definition of entropy can, like classical Boltzmann entropy, describe quantum systems becoming disordered within a chosen coarse-grained description -- the quantum mechanical equivalent of ``Spilling coffee on the table.'' This entropy generically increases, even in an isolated quantum system. However unlike classical Boltzmann entropy, it exhibits purely quantum features such as non-locality and non-commutativity. In this formalism, coarse-graining can be viewed as a sequence of measurements. And since these measurements can be chosen freely by an observer (with the aim to describe a particular physical scenario), we call this formulation {\em Observational entropy}. 

We identify two methods of coarse-graining, and thus two entropy quantities, that are particularly interesting. The first entropy can be understood as uncertainty in outcomes of two consecutive measurements, first in measuring local particle numbers and then total energy; second can be understood as uncertainty in measuring local energies. These entropies describe regions of space trying to equilibrate with each other. Both of these entropies converge to thermodynamic entropy as the system thermalizes, and they extend well to non-equilibrium situations, making them suitable candidates for the dynamical description of equilibration of isolated quantum systems.

Observational entropy elucidates the dynamics of a variety of quantum thermodynamic systems and may shed light on thorny questions such as the entropy of black holes and horizons in general, or the arrow of time in the Universe as a whole. Experimentally, it could have applications in cold atoms, where experiments on isolated quantum systems are now becoming feasible~\cite{kaufman2016quantum,trotzky2012probing}.

\vskip0.1in
We start by considering making a single observation on a quantum system characterized by
a density matrix $\R$. In analogy to classical physics, we define measurements
of the system that partition it into coarse-grained {\em macrostates}. We do this through
a set of trace-preserving projectors $\{\P_i\}_i$, indexed by $i$, acting on a Hilbert space $\HS$. 

For example, given a system with $N$ indistinguishable particles, we can coarse grain them into $p$ bins, each of width $\delta$. We wish to make observations that will give us the bin that every particle is in. To do this, we denote the particle positions by $\vec{x}=(x^{(1)},\dots,x^{(N)})$, where each element can take one of the equidistant values $x_1,\dots,x_p$. Because the particles are indistinguishable, any permutation $\pi$ of elements of $\vec{x}$ constitutes the same vector, $\vec{x}\equiv\pi(x^{(1)},\dots,x^{(N)})$. With $i\rightarrow{\vec x}$, we define a coarse-graining as a set of projectors
\[
\label{eq:x_projectors} 
\C_X=\{\P_{\vec x}^{(\delta)}\}_{\vec x}, ~ {\rm where} ~ \P_{\vec{x}}^{(\delta)}=\sum_{\vec{\tilde{x}} \in C_{\vec{x}}} \pro{\vec{\tilde{x}}}{\vec{\tilde{x}}}
\]
and $C_{\vec{x}}$ represents a hypercube of dimension $N$ and width $\delta=x_{j+1}-x_j$, that represents the possible particle positions in a single macrostate. Our coarse-graining $\C_X$ then represents measurements that can be done that will characterize the system positional macrostate at a scale of $\delta$. The above coarse-graining is written in a rigorous but fairly complicated way, but since we consider indistinguishable particles, its meaning is quite simple: it corresponds to measuring number of particles in each bin of size $\delta$.

Performing the above coarse-grained measurement does not give the precise position of the particles. After the measurement, if the particles were confined to a lattice, a further measurement could be done that would give the positional basis states $\vec{x}$ precisely. In more generality, after performing a coarse-grained measurement defined by a set of projectors $\{\P_i\}$, the number of possible outcomes of a second measurement that would determine the basis state of the system is $\tr[\P_i]$ and so with no more information, we would then assign equal weights to these different outcomes. The probability of finding the system in a particular subspace $\HS_i$ of the total Hilbert space is equal to $p_i=\tr[\P_i\R]$. Therefore the probability of finding the system in any of the basis states is $p_i/\tr[\P_i]$.

This allows us to define {\em Observational entropy} for coarse-graining $\C=\{\P_i\}$ as the Shannon entropy of these probabilities,
\[
\label{eq:def_obs_entropy}
S_{O(\C)}(\R)\equiv -\sum_{i}p_i\ln \frac{p_i}{\tr[\P_i]}.
\]
$p_{i}$ can be interpreted as a probability of a microstate of the system (described by a density matrix $\R$) to be in macrostate ``$i$'', while $V_i\equiv\tr[\P_i]$ denotes volume of that macrostate. $V_i$ is always positive and $\sum_iV_i=\dim \HS$. The above formula can be also rewritten as $S_{O(\C)}(\R)= -\sum_{i}p_i\ln p_i+\sum_{i}p_i\ln V_i$. The first part corresponds to the mean uncertainty in to which macrostate the state of the system belongs to, while the second part corresponds to the mean uncertainty about the system after the coarse-grained measurement is performed.

The idea of coarse-grained projections is mentioned very early on by von  Neumann~\cite{von2010proof} with an expression similar to this for the particular case of coarse-grained energies, that he attributes to Eugene Wigner. The general form of Eq.~\eqref{eq:def_obs_entropy} is mentioned later by Werhl~\cite{wehrl1978general}, in connection with developing a quantum mechanical master equation, and by Brun and Hartle in connection with coarse-grained histories~\cite{brun1999entropy}. By itself, it does not connect to thermodynamic entropy, for which it is necessary to consider multiple coarse-grainings, as we will do later. However it has a number of interesting properties that we studied and that are briefly discussed below. The detailed definitions and proofs are published in~\cite{SafranekDeutschAguirreObservationalEntropyLongPaper}.
\begin{itemize}
\item Observational entropy is a quantum analog of Boltzmann entropy: for a density matrix contained in a subspace $\HS_i$, i.e., $\P_i\R\P_i=\R$, its value is equal to the logarithm of the volume of the subspace,
\[
S_{O(\C)}(\R)=\ln V_i=\ln \mathrm{dim}\HS_i.
\]
This aligns with the coarse-graining interpretation that we gave: because the  basis state is not measured to more precision than the one given by coarse-graining $\C$, this entropy represents the inability of such measurements to acquire more accurate information even if the state of $\R$ is known to more precision.
\item
In classical thermodynamics, a point in phase-space belongs to a single macrostate. Due to the superposition in quantum mechanics, even a pure state can span over several macrostates. This leads to a necessity of considering non-trivial distributions $p_i$ in Eq.~\eqref{eq:def_obs_entropy}, which is where Observational entropy differs from Boltzmann entropy.
\item
The degree to which coarse-grained measurements specify a system can be made more precise by considering two coarse-grainings $\C_1$ and $\C_2$, and saying that $\C_2$ is ``finer than'' $\C_1$, denoted by writing $\C_1 \hookrightarrow \C_2$, if projectors in $\C_1$ can always be written as the sum of projectors in $C_2$. In this case, it can be proven that 
\[
S_{O(\C_1)}(\R)\geq S_{O(\C_2)}(\R).
\]
This intuitively means that entropies will be larger when the coarse-graining is coarser.
\item There are general bounds that we have proven for it:
\[
S_{V\!N}(\R)\leq S_{O(\C)}(\R)\leq \ln \mathrm{dim}\HS,
\]
where $S_{V\!N}$ is the von Neumann entropy.
\item Observational entropy is extensive. Consider a composite of $m$ sub-systems
characterized together by a separable state $\R=\R^{(1)}\otimes\cdots\otimes\R^{(m)}$.
If we impose a coarse-graining $\C=\C^{(1)}\otimes\dots\otimes \C^{(m)}=\{\P_{i_1}\otimes\dots\otimes\P_{i_m}\}_{i_1,\dots,i_m}$,
which coarse grains the different subsystems separately, then 
\[
S_{O(\C)}(\R)=\sum_{k=1}^mS_{O(\C^{(k)})}\left(\R^{(k)}\right).
\]
\item If the coarse-graining $\C$ is composed of projectors that commute with the Hamiltonian,
the Observational entropy $S_{O(\C)}(\R_t)$ (of the time-evolving density matrix $\R_t$), 
does not vary in time. 
\item Otherwise, for a large class of nonequilibrium initial states, the Observational entropy increases. A provable result is that starting with an initial state that is contained in one of the subspaces ${\cal{H}}_i$ the Observational entropy increases or remains the same,  at least for a short time.
\end{itemize}

This definition of entropy can partition Hilbert space using a single coarse-graining $\C$ that corresponds to a single measurement, for example of position, that one can
perform on the system. But to get a useful generalization of coarse-grained classical
entropies, we should consider a second coarse-graining corresponding, for example, to measurement in energy. Indeed our classical notion of a coarse-grained phase space requires consideration of two types of measurements, for example position and momentum, that in the quantum mechanical case do not commute. Therefore
we need to generalize the above definition of entropy to allow for the series
of possibly non-commuting measurements. We will find that this leads to
a surprisingly simple prescription for coarse-grained but fully quantum
mechanical entropy.

For simplicity, consider two different coarse-grainings $\C_1=\{\P_{i_1}\}_{i_1}$ and $\C_2=\{\P_{i_2}\}_{i_2}$ that may not commute. $p_{i_1i_2}=\tr[\P_{i_2}\P_{i_1}\R\P_{i_1}\P_{i_2}]$ represents the probability of obtaining result $i_1$ in the first measurement while obtaining result $i_2$ in the second measurement when two consequent measurements in bases $\C_1$ and $\C_2$ are performed on the state described by the density matrix $\R$. Equivalently, $p_{i_1i_2}$ can be interpreted as a probability of a microstate of the system (described by a density matrix $\R$) to be in a multi-macrostate $\boldsymbol{i}=(i_1,i_2)$ of volume $V_{i_1,i_2}\equiv \tr[\P_{i_2}\P_{i_1}\P_{i_2}]$. $V_{i_1,i_2}$ is always positive and $\sum_{i_1,i_2}V_{i_1,i_2}=\dim\HS$. This can be generalized further~\footnote{This is not a unique generalization, but is selected for its desirable properties; see~\cite{SafranekDeutschAguirreObservationalEntropyLongPaper}.} to give

\begin{definition}\label{def:o_entropy_general}
Let $(\C_1,\dots,\C_n)$ be an ordered set of coarse-grainings. We define \emph{the Observational entropy with coarse-grainings $(\C_1,\dots,\C_n)$} as
\[
S_{O(\C_1,\dots,\C_n)}(\R)\equiv-\!\!\sum_{i_1,\dots,i_n}\!\!p_{i_1,\dots,i_n}\ln \frac{p_{i_1,\dots,i_n}}{V_{i_1,\dots,i_n}},
\]
where the sum goes over elements such that
$p_{i_1,\dots,i_n}\equiv\tr\big[\P_{i_n}\cdots\P_{i_1}\R\P_{i_1}\cdots\P_{i_n}\big]\neq 0$, and $V_{i_1,\dots,i_n}=\tr[\P_{i_n}\cdots\P_{i_1}\cdots\P_{i_n}]$.
\end{definition}
It is important to note that in the above definition, the order of coarse-grainings \emph{does} matter: generally $S_{O(\C_1,\C_2)}\neq S_{O(\C_2,\C_1)}$. This non-commutativity is another point where Observational entropy differs from Boltzmann entropy.

For finite-dimensional systems, Observational entropy can be expressed using Kullback-Leibler divergence as
\[\label{eq:KL_divergence_entropy}
S_{O(\C_1,\dots,\C_n)}(\R)=\ln \dim \HS - D_{KL}\big(P(\R)\big|\big|P\left(\R_{\mathrm{id}}\right)\!\!\big),
\]
where $P_{i_1,\dots,i_n}(\R)=\tr\big[\P_{i_n}\cdots\P_{i_1}\R\P_{i_1}\cdots\P_{i_n}\big]$.
The Observational entropy therefore measures the distance between probability distributions of measurement outcomes produced by the density matrix $\R$ and by the maximally-uncertain density matrix $\R_{\mathrm{id}}=\hat{I}/\dim \HS$.

We can generalize the notion of finer coarse-grainings to multiple coarse-grainings and prove the following theorem. (For details see~\cite{SafranekDeutschAguirreObservationalEntropyLongPaper}.)
\begin{theorem}
\label{thm:bounded_multiple}
For any ordered set of coarse-grainings $(\C_1,\dots,\C_n)$ and any density matrix $\R$,
\begin{align}
S_{V\!N}(\R)\leq S_{O(\C_1,\dots,\C_n)}(\R)\leq \ln \mathrm{dim}\HS,\\
S_{O(\C_1,\dots,\C_n)}(\R)\leq S_{O(\C_1,\dots,\C_{n-1})}(\R).
\end{align}
$S_{V\!N}(\R)= S_{O(\C_1,\dots,\C_n)}(\R)$ if and only if for all $i_1,\dots,i_n$ there exists $\P_\rho\in\C_\R$ such that $\P_{i_n}\cdots\P_{i_1}\P_\rho=\P_{i_n}\cdots\P_{i_1}$, $\P_{i_k}\in \C_k$. $S_O(\R)=\ln \mathrm{dim}\HS$ if and only if for all $i_1,\dots,i_n$, $p_{i_1,\dots,i_n}=\frac{V_{i_1,\dots,i_n}}{\mathrm{dim}\HS}$. $S_{O(\C_1,\dots,\C_n)}(\R)= S_{O(\C_1,\dots,\C_{n-1})}(\R)$ if and only if for all $i_1,\dots,i_{n}$, $p_{i_1,\dots,i_{n}}=\frac{V_{i_1,\dots,i_{n}}}{V_{i_1,\dots,i_{n-1}}}p_{i_1,\dots,i_{n-1}}$.
\end{theorem}
In the above, we used coarse-graining given by the density matrix $\C_\R$. For a Hermitian operator $\hat{A}$, $\C_{\hat{A}}$ consists of projectors from the spectral decomposition of $\hat{A}$.

With general Definition~\ref{def:o_entropy_general} in mind, it is possible to consider many possible kinds of Observational entropies, by considering different types of composite coarse-grainings defined in terms of sequences of coarse-grained measurements. It is not obvious that any of these have any relation to thermodynamic entropy, but we now describe two versions that do bear a close connection.

In Eq.~\eqref{eq:x_projectors} we introduced coarse-graining in position space with $p$ number of bins. Consider these and ``fine-grained" energy projectors 
\[
\C_E=\{\P_E\}_E,\quad \P_E= \pro{E}{E}.
\]
We construct entropy 
\[
S_{xE} \equiv S_{O(\C_X,\C_E)}(\R),
\]
which corresponds to measuring the coarse-grained position of the system (or equivalently, measuring the local particle numbers), and then its energy.

The second entropy is similar in spirit but it employs a different coarse-graining. We start by considering Hilbert space divided into two parts $\HS^{(1)}$ and $\HS^{(2)}$,
the joint system being $\HS=\HS^{(1)}\otimes\HS^{(2)}$.
The Hamiltonian $\hat{H}$ can then be separated into three terms
\[
\label{eq:H=H1+H2+Hint}
\hat{H}=\hat{H}^{(1)}\otimes \hat{I}+\hat{I}\otimes\hat{H}^{(2)}+\epsilon\hat{H}^{(\mathrm{int})},
\]
where $\hat{H}^{(1)}$ and $\hat{H}^{(2)}$ are the Hamiltonians that describe internal interactions in the first and second systems respectively, and $\hat{H}^{(\mathrm{int})}$ is an interaction term. For large subsystems and local interactions, the magnitude of this term is expected to be small and hence we have introduced a parameter $\epsilon$ to indicate this. Consider a coarse-graining that projects to the eigenstates of the local Hamiltonians $\hat{H}^{(1)}$ and $\hat{H}^{(2)}$, which corresponds to simultaneous measurements of local energies.
We call this the \emph{factorized Observational entropy} (FOE). It can be formally written as 
\[\label{def:non-equilibrium_entropy}
S_{FOE}\equiv S_{O{\displaystyle (}\C_{\hat{H}^{(1)}}\otimes \C_{\hat{H}^{(2)}}{\displaystyle )}}(\R).
\]
This can be easily generalized to an arbitrary number $m$ of local Hamiltonians,
rather than two.

\begin{figure}[t]
\includegraphics[width=1\hsize]{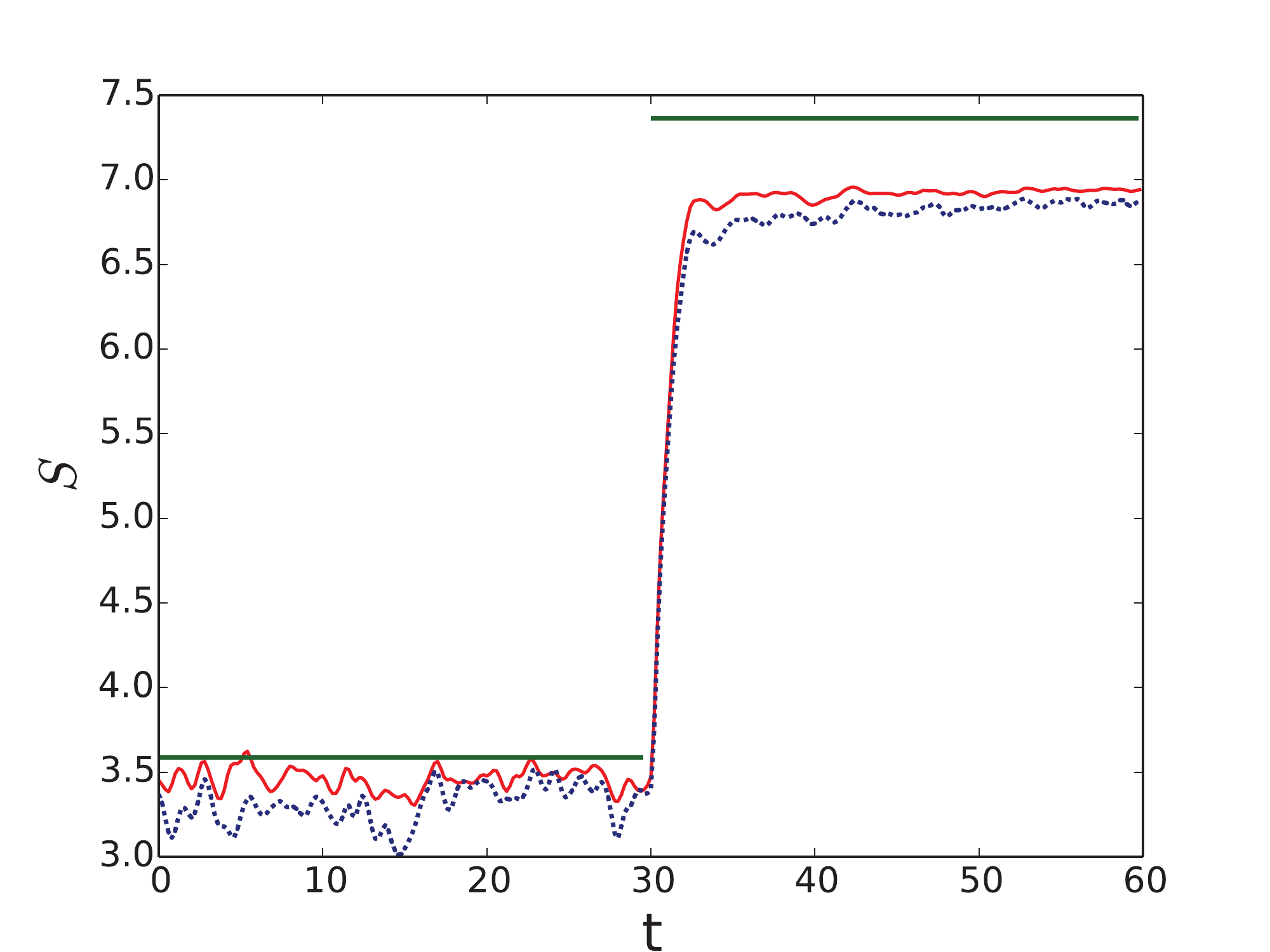}
\caption
{
Time evolution of $S_{xE}$ (line) and the factorized Observational entropy FOE (dashed line) starting in a pure thermal state. After $t=30$, the right wall is expanded to double the system size and the system continues to evolve. The straight lines represent thermodynamic entropies of the canonical ensemble. This graph shows that for a typical state, $S_{xE}$ and FOE models the dynamical process of equilibriation between the two regions.
}
\label{fig:dynamics}
\end{figure}

While mathematially distinct, we will see below that $S_{xE}$ and FOE have similar behavior, and  can be interpreted in a similar fashion. 
Both measure how democratically the total energy is distributed over the regions of space, which are defined by coarse-graining $\C_X$, Eq.~\eqref{eq:x_projectors}, in the case of $S_{xE}$, and by separation into local Hamiltonians, Eq.~\eqref{eq:H=H1+H2+Hint}, in the case of FOE. Both entropies are maximal when the energy contained in each region is roughly proportional to the size of the region, and small when energy is unevenly distributed -- for example when a small region contains a large amount of energy while a large region contains a small amount of energy. These entropies therefore describe how close these regions are to thermal equilibrium with each other. Entropy increase then signifies regions equilibrating with each other by exchanging heat until they attain the same temperature, at which point both $S_{xE}$ and FOE achieve the thermodynamic entropy of the full system.

\begin{figure}[t]
\includegraphics[width=1\hsize]{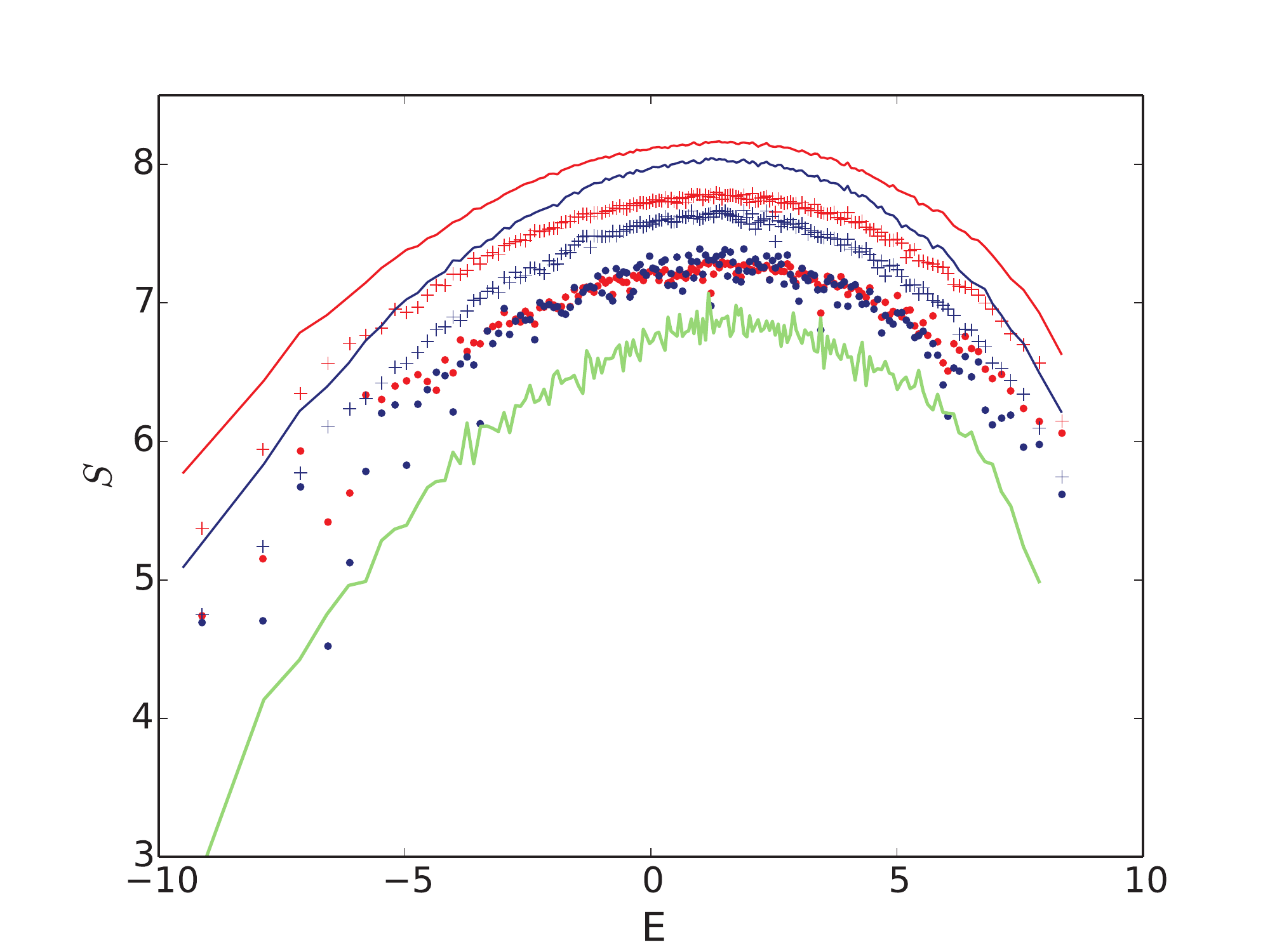}
\caption
{
The top curves show Observational entropies $S_{xE}$ (light red) and FOE (dark blue) for a microcanonical state (line), a random superposition of neighboring energy eigenstates (crosses), and energy eigenstates (dots), from top to bottom. The lowest curve is the microcanonical entropy given by logarithm of the density of states. Both $S_{xE}$ and FOE of random superposition of neighboring energy eigenstates approximate the thermodynamic entropy. Since these states model typical states of an isolated quantum system in far future, this graph provides an extensive numerical evidence that for most initial states, $S_{xE}$ and FOE converge to thermodynamic entropy, even for genuinely closed quantum systems.
}
\label{fig:SXEvsE}
\end{figure}

Let us start with a numerical analysis of these quantities. We consider a one dimensional lattice model of spinless fermions, with both nearest-neighbor (NN) and next-nearest-neighbor (NNN) hopping and interactions~\cite{santos2010onset} of strengths $V$ and $V'$ respectively. We always take $\hbar=V=t=1$. For generic systems we choose the well-studied case $U'=t'=0.96$~\cite{santos2010onset,deutsch2013microscopic,santos2012weak}. We employ hard wall boundary conditions for our numerical experiments so that we can study the expansion of a gas from a smaller to a larger box.

First we investigate the dynamics of the two Observational entropies $S_{xE}$ and FOE, both of which  are coarse-grained into four subsystems ($p=m=4$) of length 4 in the full system of length $L=16$. The graph of the evolution is shown in Fig.~\ref{fig:dynamics}. We start with a system with $N=4$ particles confined to a box of size $L=8$. The system starts in what can be described as a ``pure thermal state.'' It is a superposition of all energy eigenstates, each eigenstate having a random complex amplitude drawn from a distribution with
a variance given by the Gibbs distribution at inverse temperature
$\beta=1$. For $t < 30$ the system is in equilibrium.
At $t=30$, we suddenly enlarge the box to size $L=16$ and compute the continued evolution. Both entropies increase rapidly but smoothly, until they reach equilibrium. The dashed lines represent entropy of the canonical distribution. Because of finite-size effects, this differs from the computed values of the $S_{xE}$ and $S_{F}$ by approximately $10\%$. This behavior is robust, and holds over a wide variety of initial states we have investigated~\cite{SafranekDeutschAguirreObservationalEntropyLongPaper}. We also analyzed the integrable case, $U'=t'=0$. As expected, the integrable case shows substantially larger fluctuations~\cite{SafranekDeutschAguirreObservationalEntropyLongPaper}.

To investigate behavior of these two entropies in more detail, we also plot $S_{xE}$ and FOE as functions of energy for various equilibrium states as shown in Fig.~\ref{fig:SXEvsE}; this is particularly relevant for studying the long-time limit. Both entropies are coarse-grained into 4 subsystems ($p=m=4$) of the full generic (non-integrable) system of size $L=20$, and computed for energy eigenstates, random superposed pure states, and microcanonical
mixed states. The random superposed pure states were obtained by superposing $k=30$ neighboring energy eigenstates with complex amplitudes drawn uniformly from the unit disk, then normalizing. The microcanonical states were obtained by adding together 
the density matrices of $k=30$ neighboring energy eigenstates with equal weights. Because of significant finite size effects, we eschew using the canonical ensemble for comparison, and instead focus on the microcanoncial ensemble given by the density of states $\rho(E)$; we plot $S_{DOS} \equiv \ln(\rho(E) \Delta E)$~\footnote{To give it the right units, we have multiplied $\rho$ by a typical energy $\Delta E = \sigma(E)/\sqrt{N}$, where $\sigma$ computes the standard deviation and $N$ is the number of particles. This choice of $\Delta E$ is rather arbitrary as it is unimportant in the thermodynamic limit.}. $S_{DOS}$ gives an entropy that, up to an unimportant additive constant, is in thermodynamic limit equivalent to the thermodynamic entropy given by the canonical ensemble~\cite{ruelle1999statistical}.

The results for the two quantities are quite similar; the same is true for the time-dependent analysis shown in Fig.~\ref{fig:dynamics}.   It can be shown that in the $\epsilon \rightarrow 0$ limit in Eq.~\eqref{eq:H=H1+H2+Hint}, $S_{xE}$ and FOE are the same, and there are strong arguments that the quantities are very closely tied for finite $\epsilon$ (see~\cite{SafranekDeutschAguirreObservationalEntropyLongPaper}).     

As shown on Fig.~\ref{fig:dynamics}, both $S_{xE}$ and FOE approximate the thermodynamic entropy in the long-time limit. Figure~\ref{fig:SXEvsE} provide even more compelling evidence for this convergence, as follows.

It is possible to prove that up to order $\epsilon$, FOE of a canonical state is equivalent to the canonical entropy~\cite{SafranekDeutschAguirreObservationalEntropyLongPaper}, which is equivalent to microcanonical entropy in the thermodynamic limit~\cite{ruelle1999statistical}. Curves for both $S_{xE}$ and FOE approximate the microcanonical entropy computed from the density of states -- in fact, they are almost parallel to each other. The differences of order $O(1)$ are unimportant in the thermodynamic limit. The superposed states have random phases, meaning that they describe the state of a typical wave-function at some time far in the future, which provides additional support to the claim that in the long time limit, and for generic systems, these two Observational entropies converge to the thermodynamic entropy.

Convergence of $S_{xE}$ and FOE to the thermodynamic entropy can be also shown analytically~\cite{SafranekDeutschAguirreObservationalEntropyLongPaper} for generic (i.e., non-integrable) systems of large size, by using connections between non-integrable systems and random matrix theory. These results show that both Observational entropies, in the form of $S_{xE}$ and FOE, extend the idea of classical Boltzmann entropy to quantum mechanical systems.

It is worthwhile briefly comparing the above approach with other well-known entropies used for closed quantum systems. The entanglement entropy is also closely related to the thermodynamic entropy in equilibrium~\cite{deutsch2010thermodynamic,deutsch2013microscopic,santos2012weak}. But it is a distinct quantity that is fundamentally different from $S_{xE}$ or FOE. For example, if the state is a product state, then the entanglement entropy is zero, but $S_{xE}$ is not. Thermodynamic entropy of the complete system should still be large, and thus the entanglement entropy cannot give us a sensible measure, at least in this case, for the thermodynamic entropy. On the other hand $S_{xE}$ is largely unaffected by this lack of entanglement for short ranged systems. The diagonal entropy~\cite{tolman1938principles,ter1954elements,jaynes1957information2,polkovnikov2011microscopic}, can be defined as Observational entropy with coarse-graining given by (non-degenerate) Hamiltonian $\hat{H}$, as $S_{\mathrm{diag}}\equiv S_{O(\C_{\hat{H}})}$. This quantity stays constant in an isolated system, unless one allows transitions between instantaneous energy levels~\cite{polkovnikov2011microscopic}, or external operations on the system~\cite{ikeda2015second}. On the other hand, both $S_{xE}$ and FOE rise even in a genuinely isolated system.

Observational entropy may play a useful role in experiments, for example on cold atoms, in which these kinds of measurements and coarse-grainings are possible. It is hard to measure the  entanglement entropy between two subsystems directly~\cite{rispoli2016measuring}, and to compute it one needs to know the full density matrix for at least one of the subsystems, which requires a very large set of measurements. On the other hand, to obtain $S_{xE}$, we first determine the coarse-grained position of particles. This is equivalent to measuring the coarse-grained density, which is frequently performed in cold atom experiments~\cite{levin2012ultracold,kaufman2016quantum}. Then the state energy is observed~\cite{Villa2018cavityassisted}. Even if the apparatus is not precise enough to distinguish individual eigenstates, the Observational entropy with finite energy coarse-graining can still be calculated theoretically, and compared with experimental data.

\vskip0.1in
We have argued through both analytical and numerical work that it is indeed possible to extend coarse grained entropy to quantum mechanics, and shown that for a variety of initial states and for non-integrable systems,  this entropy generically rises, approaching the correct thermodynamic value. It is easily understood in terms of performing subsequent measurements, has the mathematical properties expected of entropy, and has close ties to experimental techniques. Thus Observational entropy is a very promising candidate for understanding the non-equilibrium evolution of entropy, and the second law of thermodynamics, in closed quantum systems.

\begin{acknowledgements}
This research was supported by the Foundational Questions Institute (FQXi.org), of which AA is Associate Director, and by the Faggin Presidential Chair Fund.
\end{acknowledgements}

\bibliographystyle{apsrev4-1}
\bibliography{observational_entropy}

\end{document}